\documentclass[preprint,journal]{vgtc}              % review (journal style)
\usepackage{amsmath}

\onlineid{1033} 

\preprinttext{Submitted to IEEE Vis 2024}

\vgtccategory{Research}

\vgtcpapertype{application/design study}

\title{ClusterRadar: an Interactive Web-Tool for the Multi-Method Exploration of Spatial Clusters Over Time}

\author{%
  \authororcid{Lee Mason}{0000-0002-7814-099X},
  \authororcid{Blánaid Hicks}{0000-0002-5730-9469}, and 
  \authororcid{Jonas S. Almeida}{0000-0002-7883-7922}
}

\authorfooter{
  \item
  	Lee Mason is with Queen's University Belfast and the National Cancer Institute.
  	E-mail: masonlk@nih.gov
  \item
  	Blánaid Hicks is with Queen's University Belfast.
  	E-mail: b.hicks@qub.ac.uk.

  \item Jonas Almeida is with the National Cancer Institute.
  	E-mail: jonas.dealmeida@nih.gov
}

\abstract{%
Spatial cluster analysis, the detection of localized patterns of similarity in geospatial data, has a wide-range of applications for scientific discovery and practical decision making. One way to detect spatial clusters is by using local indicators of spatial association, such as Local Moran's I or Getis-Ord Gi*. However, different indicators tend to produce substantially different results due to their distinct operational characteristics. Choosing a suitable method or comparing results from multiple methods is a complex task. Furthermore, spatial clusters are dynamic and it is often useful to track their evolution over time, which adds an additional layer of complexity. ClusterRadar is a web-tool designed to address these analytical challenges. The tool allows users to easily perform spatial clustering and analyze the results in an interactive environment, uniquely prioritizing temporal analysis and the comparison of multiple methods. The tool's interactive dashboard presents several visualizations, each offering a distinct perspective of the temporal and methodological aspects of the spatial clustering results. ClusterRadar has several features designed to maximize its utility to a broad user-base, including support for various geospatial formats, and a fully in-browser execution environment to preserve the privacy of sensitive data. Feedback from a varied set of researchers suggests ClusterRadar's potential for enhancing the temporal analysis of spatial clusters.  

}


\teaser{
  \centering
  \includegraphics[width=\linewidth, alt={A screenshot of the ClusterRadar tool.}]{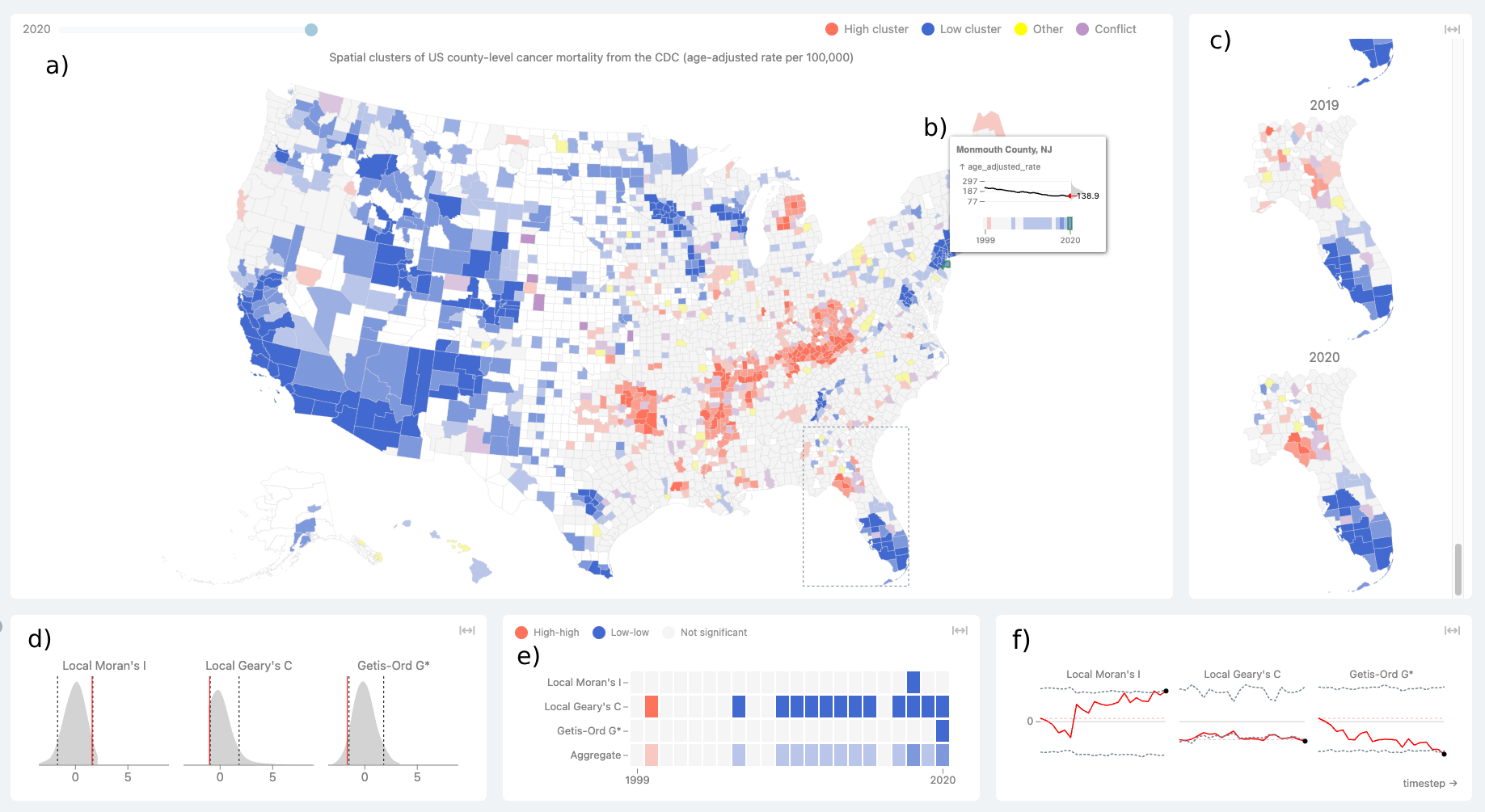}
  \caption{
The main dashboard of the ClusterRadar tool, showing the various plot panels and the graphical tooltip. The dashboard is divided into two rows of plots: the panels in the top row show the geospatial distribution of the cluster assignments, and the panels in the bottom row show various additional details about the results. The five main panels are: a) the primary cluster map panel, c) the zoomed map reel panel, d) the statistical density plot panel, e) the cluster assignments cell plot panel, and f) the statistical time-series plot panel. 
In this screenshot, the user has hovered over a location (Monmouth County, NJ) showing b) the graphical tooltip, which gives additional information about that location's data and cluster assignments over time. Because the user is hovering over a specific location, the bottom row plots show results related to that location. The user has also selected a sub-region around Florida, and that region is consequently shown in the zoomed map reel panel. The data shown in this screenshot is yearly US county-level age-adjusted cancer mortality data from the CDC, from 1999-2020. ClusterRadar can be accessed at \href{https://episphere.github.io/ClusterRadar}{https://episphere.github.io/ClusterRadar}
  }
  \label{fig:teaser}
}

\renewcommand{\manuscriptnotetxt}{}

\graphicspath{{figs/}{figures/}{pictures/}{images/}{./}} % where to search for the images

\usepackage{tabu}                      % only used for the table example
\usepackage{booktabs}                  % only used for the table example
\usepackage{lipsum}                    % used to generate placeholder text
\usepackage{mwe}                       % used to generate placeholder figures

\usepackage{mathptmx}                  % use matching math font

\begin{document}

%%%%%%%%%%%%%%%%%%%%%%%%%%%%%%%%%%%%%%%%%%%%%%%%%%%%%%%%%%%%%%%%
%%%%%%%%%%%%%%%%%%%%%% START OF THE PAPER %%%%%%%%%%%%%%%%%%%%%%
%%%%%%%%%%%%%%%%%%%%%%%%%%%%%%%%%%%%%%%%%%%%%%%%%%%%%%%%%%%%%%%%

%\firstsection{Introduction}

% TODO: Make sure all figures are referenced in the text!
% TODO: Put through some sort of more advanced spell check

\maketitle

\section{Introduction}

% ***
Space plays a critical role in many real-world phenomena, where the proximity of entities in a system often effects the strength and nature of interactions
\cite{unwin1996gis,weeks2004role,tobler1970computer}. % CITE DELETE, USE OTHER REFERENCES FROM PAPER
This idea is succinctly captured in Tobler's first law of geography, which states that "everything is related to everything else, but near things are more related than distant things" \cite{tobler1970computer,miller2004tobler}. The field of geospatial analysis offers a powerful set of methods for understanding these relationships \cite{park_understanding_2021, auchincloss_review_2012}. Techniques like spatial clustering, machine learning, spatial statistics, and network analysis allow analysts to uncover informative relationships within geospatial data \cite{ripley2005spatial, grubesic2014spatial, abd2018geospatial}. The growing availability of geospatial data, coupled with an increasingly interconnected world, mean that sophisticated geospatial techniques are increasingly important in driving scientific discovery and supporting informed decision making \cite{grubesic2014spatial,auchincloss_review_2012}. 
% **

One valuable application of spatial analysis is the detection of spatial clusters: groupings of neighboring geospatial features which exhibit significant similarity across certain attributes \cite{grubesic2014spatial, varghese2013spatial}. For example, a spatial cluster might be a collection of neighboring counties with similar rates of breast cancer. Spatial cluster detection is a widely employed technique in geospatial analysis, crucial for advancing scientific understanding of complex systems and tackling practical challenges \cite{kulldorff1995spatial, varghese2013spatial, grubesic2014spatial}. Spatial clustering has been extensively applied across disciplines, including cancer cluster detection in epidemiology \cite{jacquez2003local}, crime cluster analysis in criminology \cite{quick2013exploring}, and the study of weather events in climatology \cite{lyra2014cluster}. Robust spatial cluster detection methodologies, particularly those founded on statistical rigor, empower analysts to expand beyond simple visual pattern recognition in maps \cite{amin2019spatial}.

% **
The field of spatial cluster analysis continues to evolve, addressing the expanding analytical requirements of researchers across an increasingly diverse set of disciplines \cite{ahasan2020applications, park_understanding_2021,auchincloss_review_2012}. There is a growing need to expand the scope of spatial clustering methods, including the deeper integration of temporal data, and better support for the comparison of results across multiple methods. \cite{auchincloss_review_2012, grubesic2014spatial, lin2015lung, hanson2002alcohol}. Despite the potential benefits, tools which directly support these advancements remain limited and often require specialized knowledge, a significant problem as interest in spatial analysis continues to grow among individuals with a diverse set of skillets \cite{ahasan2020applications}.

% *
Here, we introduce ClusterRadar, a web-tool designed to meet the growing need for a user-friendly environment in which to analyze spatial clusters over time and across multiple methods. ClusterRadar runs fully on the client-side in the user's browser, avoiding the need for installation and preserving the privacy of the user's data. The core feature of the tool is an interactive dashboard consisting of five panels, each providing a different perspective on the methodological and temporal aspects of multi-method, spatiotemporal clustering results. In this paper, we provide details of the methods used in the tool, describe the design of the dashboard, discuss the tool's implementation, and provide an analysis of initial feedback from researchers with varied backgrounds. The main contributions of this work are:
\begin{itemize}
  \item ClusterRadar: an interactive web-tool that allows users to perform spatial clustering and explore the results in an interactive dashboard, with a unique focus on the comparison of results across multiple methods and from multiple time points. ClusterRadar can be accessed at \href{https://episphere.github.io/ClusterRadar}{https://episphere.github.io/ClusterRadar}
  \item A set of interactive visual elements designed to help an analyst interpret multi-method, longitudinal spatial clustering results. Each of these elements is designed to show a different aspect of the results, making effective interpretation more accessible to a diverse set of users. 
  \item An analysis of feedback for ClusterRadar from multiple users of varied backgrounds. This feedback could prove valuable in the design of future analytical environments seeking to expand the scope of spatial clustering analysis. 
\end{itemize}

\section{Background overview}

\subsection{Geospatial data}
Geospatial data comprises data points explicitly linked to geographic locations on the Earth's surface. These references are typically given as points (representing precise coordinates), lines (representing paths or routes), or areas (polygonal regions) \cite{kanankege_introductory_2020}. Geospatial data can employ various coordinate systems, such as latitude and longitude, to define spatial locations. Areal data, such as countries or census regions, is highly prevalent because the often standardized and familiar area definitions facilitate integration across multiple data sources \cite{banerjee2016spatial}. While less common, line data (e.g., roads) sometimes holds analytical value. Point data can often be aggregated into areas, which can simplify analysis, improve statistical power, or facilitate linkage with other datasets \cite{kanankege_introductory_2020}. 

\subsection{Spatial cluster analysis}

Spatial clusters are defined in a number of ways, depending on the type of geospatial data and the specific analytical goals being tackled. Broadly, a spatial cluster is a grouping of geographically related features that exhibit a substantial degree of concentration or similarity \cite{grubesic2014spatial, varghese2013spatial}. There are many different approaches to spatial clustering, including partition clustering, hierarchical clustering, density-based clustering, and LISA-based clustering \cite{varghese2013spatial, grubesic2014spatial}. LISA-based clustering methods use local indicators of spatial association (LISAs) to determine whether each location belongs to a spatial cluster \cite{moraga2011detection, anselin1995local}. They are popular due to their robust statistical foundation, interpretability, and wide-spread implementation in different analytical environments \cite{grubesic2014spatial, bivand2018comparing}. 

\subsection{Spatial autocorrelation}

Spatial autocorrelation is a fundamental concept in spatial analysis describing the extent to which geographically proximate locations exhibit similar attributes \cite{getis2009spatial, anselin1995local}. Spatial autocorrelation can be quantified using several different statistics, the most prominent being Moran's I and Geary's C \cite{moran1950notes, geary1954contiguity}. Both of these statistics indicate whether or not a dataset exhibits a greater degree of spatial autocorrelation than would be expected by random chance, but they take different numerical approaches to do so. Spatial autocorrelation can be positive, meaning that locations which are near each other tend to exhibit similar attribute values, or negative, meaning that locations which are near each other tend to exhibit dissimilar attribute values.
%Many real-world datasets exhibit positive spatial autocorrelation; negative spatial autocorrelation is much less common \cite{griffith2019negative}. 

Global spatial autocorrelation statistics  are only capable of indicating whether or not spatial autocorrelation is present in a dataset but not at which locations it occurs \cite{anselin1995local}. In reality, geospatial datasets often exhibit pockets of both positive and negative spatial autocorrelation, alongside regions which don't show any substantial spatial autocorrelation, and it can be useful to analyze this heterogeneity. For this purpose, local spatial autocorrelation statistics are required. Unlike global spatial autocorrelation statistics, local spatial autocorrelation statistics are a property of each location in the dataset, indicating whether a specific location is similar to its neighbors (a spatial cluster), or dissimilar to its neighbors (a spatial outlier). The majority of global spatial autocorrelation statistics have local counterparts, such as Local Moran's I and Local Geary's C \cite{anselin1995local}. Local spatial autocorrelation statistics are one (major) subset of local indicators of spatial association (LISA), which cover a broader range of local association relationships \cite{anselin1995local}.

\subsection{Local Indicators of Spatial Association (LISA)}

A local indicator of spatial association (LISA) is a statistical measure that quantifies some degree of spatial association between a particular location and its neighbors within a geospatial dataset \cite{anselin1995local}. In addition to local spatial autocorrelation statistics, a prominent example of a LISA is the Getis-Ord Gi/Gi* family of statistics, local indicators which detect whether or not a location belongs to "hot-spot" or "cold-spot". Hot-spots are statistically significant clusters of high values, while cold-spots are statistically significant clusters of low values \cite{ord1995local, getis1992analysis}. Unlike Getis-Ord Gi/Gi*, Local Moran's I and Local Geary's C are incapable of directly detecting hot-spots and cold-spots, they can only detect whether or not a location is significantly similar or dissimilar to its neighbors. On the other hand, Getis-Ord Gi/Gi* are incapable of detecting spatial outliers. While Local Moran's I and Local Geary's C are not directly capable of detecting hot-spots and cold-spots, additional analysis is often performed to distinguish hot-spots and cold-spots among areas exhibiting significant local spatial autocorrelation \cite{anselin1995local}. All of the aforementioned LISAs (Local Moran's I, Local Geary's C, and Getis-Ord Gi/Gi*) are frequently used for detecting spatial clusters, though the spatial clusters should be interpreted differently due to the different goals of the methods. For more information on the formulation and interpretation of these statistics, see \cref{sec:methods}. 

%TODO: Examples of the LISAs being applied to spatial clustering.

\section{Related Work}

\subsection{Temporal analysis of spatial clusters}

There are several ways to incorporate temporal data into the analysis of spatial clusters. One approach utilizes methods specifically designed for the detection of spatiotemporal clusters, such as space-time scan statistics, spatiotemporal models, and modified spatial statistics \cite{kulldorff1997spatial, ansari2020spatiotemporal, kisilevich2010spatio, lee2017extending, shi2019spatiotemporal}. These approaches, while potentially powerful, can be complex and their results difficult to interpret \cite{ansari2020spatiotemporal}. An alternative approach is to apply static spatial cluster detection methods to data across different time points, comparing the resulting clusters over time. This longitudinal approach allows an analyst to use the rich suite of static spatial clustering methods available, which are often easier to interpret than dedicated spatiotemporal methods and have wider support in geospatial analysis software (especially in software accessible to non-expert users). This longitudinal approach has been applied to a wide variety of problems, including analyzing temporal variations in clusters of obesity \cite{hughey2018spatial}, dengue cases \cite{saita2022temporal}, HIV/AIDS \cite{zulu2014analyzing}, healthcare inequalities \cite{dong2023measuring}, vegetation fragmentation \cite{kowe2019exploring}, and more. 

\subsection{Multi-method analysis of spatial clusters}

There are many different methods for detecting spatial clusters and they often produce vastly different results \cite{grubesic2014spatial}. Several works compare different spatial clustering methods on various geospatial datasets, both real and simulated. For example, \cite{grubesic2014spatial} compares seven different spatial clustering methods of different types, including two LISA-based methods: Local Moran's I and Getis-Ord Gi*. The paper describes the substantial differences between the methods and the difficulties in choosing an appropriate method for a given problem. Other relevant works have compared LISA-based clustering methods with scan statistic-based clustering methods \cite{hanson2002alcohol,lin2015lung}, and different LISA-based methods to each other \cite{getis1992analysis, anselin1995local,lin2020integrative,brooks2019advantages, abdulhafedh_novel_2017}. 
%Differences in results and interpretation of LISA-based clustering methods have been recognized from their inception \cite{getis1992analysis,anselin1995local}. 
Due to differences in results from different LISA-based clustering methods, many comparative works recommend that multiple methods be applied in tandem \cite{hanson2002alcohol,lin2015lung,getis1992analysis,abdulhafedh_novel_2017, kiani2023comparing}. This approach is extensively followed in the literature, with many geospatial clustering analyses applying more than one method \cite{rosenberg1999spatial, erdogan2009explorative, fahad2022developing, kowe2019exploring }. 

\subsection{Interactive spatial dashboards}

Spatial dashboards have become a popular environment in which to analyze spatial data \cite{ praharaj2023deploying, sulaiman2020geospatial,slingsby2023gridded, deng2023visualizing}. There is an extensive and growing body of work regarding these dashboards, including works which introduce novel spatial dashboards which cover specific spatial analytical techniques or specific spatial datasets \cite{ praharaj2023deploying, sulaiman2020geospatial, figgemeier2021geo}. There is also a growing body of work in the literature regarding the design of spatial dashboards \cite{yoshizumi2020review}, including works which attempt to address some of the challenging problems such as the use of color in maps \cite{ brewer1997mapping, brychtova2015exploring}, how temporal data can be integrated into spatial visualization \cite{andrienko2003exploratory, andrienko2007geovisual, pena2019comparison, slingsby2023gridded, deng2023visualizing}, and how carefully designed interactivity can improve the power of spatial environments \cite{schumann2011analytical, crampton2002interactivity, roth2015interactivity}. The COVID-19 pandemic saw a proliferation of spatial dashboards, familiarizing the concept to a wider user-base and increasing broad interest across disciplines \cite{praharaj2023deploying, usman2022multiscale, slingsby2023gridded}.

\subsection{Spatial clustering software}

Several applications and libraries support spatial cluster analysis. ArcGIS, a commercial software, offers supports for Local Moran's I and Getis-Ord Gi* \cite{scott2009spatial}. Open-source alternatives exist, such as QGIS \cite{moyroud2018introduction} (with limited LISA-based clustering via plug-ins) and CrimeStat \cite{levine2004crimestat} (specializing in crime data analysis using methods like Local Moran's I and Getis-Ord Gi/Gi*). Programming libraries like PySAL (Python) and spdep (R) support LISA-based clustering but require coding expertise \cite{rey2009pysal, bivand2005spdep}. GeoDa, an open-source tool dedicated to spatial analysis, is a strong choice for LISA-based methods \cite{anselin2009geoda}. It offers an interactive interface for analysis and visualization, but doesn't support temporal analysis or the direct comparison of different methods. SaTScan supports spatiotemporal cluster analysis but focuses on scan statistic-based methods and doesn't support LISA-based approaches \cite{block2007software}. While most tools require local installation (some with limited support for different operating systems), web-based options like GeoDa-Web and its associated JavaScript library (jsgeoda) are emerging, though currently with limited features \cite{li2015geoda}. 

\section{Design considerations}

We have formulated five key design considerations based primarily on important challenges and analytical ideals identified in the literature. 

\begin{itemize}
  \item \textbf{D1: Representation of temporal dynamics of spatial clusters.} Spatiotemporal analysis is rapidly gaining importance across diverse disciplines \cite{auchincloss_review_2012, nazia2022methods, shi2019spatiotemporal}. Historically, a major shortcoming in spatial analysis has been neglecting temporal dynamics \cite{thompson2019systematic}. However, the growing availability of user-friendly spatiotemporal software is steadily overcoming this limitation \cite{block2007software}. One valuable approach involves the longitudinal analysis of spatial clusters over time, which reveals how spatial clusters evolve \cite{tao2023applying, zulu2014analyzing, dong2023measuring}.  Despite the utility of this technique, software with dedicated support remains scarce. This highlights the need for applications that prioritize the clear representation of temporal dynamics within spatial clusters.
  \item \textbf{D2: Comparison of results from multiple spatial clustering methods.} Spatial clustering encapsulates a wide variety of approaches, often producing vastly different results \cite{grubesic2014spatial, kiani2023comparing}. There is limited theoretical guidance on how to select an appropriate method \cite{grubesic2014spatial}. Comparative studies of spatial clustering methods often recommend using multiple methods simultaneously and interpreting the results in conjunction \cite{hanson2002alcohol,lin2015lung,getis1992analysis,abdulhafedh_novel_2017, kiani2023comparing} --- a common approach across multiple disciplines. This highlights the need for applications which perform multiple spatial clustering methods at once and facilitate the comparison of results across methods.
  \item \textbf{D3: Application of varied interactive graphical elements to simplify the analysis of complex results.} Representing temporal and multivariate geospatial results can be challenging to the competing demands of the various visual elements --- this is a well-recognized problem in geospatial visualization \cite{andrienko2003exploratory, pena2019comparison, mcnabb2019multivariate, opach2014choropleth}. A multi-plot, interactive dashboard can tackle these challenges by representing different aspects of the data in different visualizations, drawing on the unique strengths of each to improve overall clarity \cite{opach2014choropleth, sack2014interactive, janes2013effective}. Interactivity further eases the complexity of this analysis, revealing details only as they are required, and helping users stay oriented while exploring the data \cite{sack2014interactive, cui2019visual}. This highlights the need for applications which employ multi-faceted, interactive graphics to facilitate the exploration of complex geospatial results.
  \item \textbf{D4: Goal focused and appropriately scoped design to ensure usability for non-expert users. } Powerful geospatial software often presents a steep learning curve for non-experts because it requires a robust understanding of complex geospatial concepts \cite{ahasan2020applications, zhu2021next}. As geospatial methods becomes increasingly integrated into diverse analytical pipelines, the need for user-friendly geospatial analytical applications grows. To make geospatial methods more accessible, a goal-driven approach is required \cite{zhu2021next}. This approach focuses on the desired analytical outcomes, rather than requiring users to understand the low-level steps required to achieve them.  While feature-rich geospatial software offers versatility, it can overwhelm non-experts with its complexity \cite{marzi2022nature}.  This highlights the need for simpler, goal-oriented tools designed for specific analytical tasks.
  \item \textbf{D5: In-browser web implementation to ensure FAIR distribution and privacy preservation.} There has been a recent emphasis in science on adherence to the FAIR principles (findability, accessiblity, interoperability, and reproducability), including for software \cite{garcia2023moving, barker2022introducing}. Due to its ubiquity, familiarity, and inherent support for information sharing, the web provides a naturally FAIR place in which to distribute software. However, server-reliant software requires the user to upload their data, which may violate privacy requirements of sensitive data (common in fields like epidemiology). This highlights the need for applications which are distributed on the web and run fully client-side inside the sandbox of the user's browser.

\end{itemize}

\section{Methods} \label{sec:methods}

% TODO: Make sure to mention the z-normalization of the statitistics

In this section, we will go into greater detail about the spatial clustering methods used in ClusterRadar.

\subsection{Normalization}

Given a list of spatially-referenced values $X = [x_1, x_2, ..., x_n]$ the first step is to z-score normalize each value, which simplifies the downstream calculations: 
\begin{equation}
z_i = \frac{(x_i - \mu)}{\sigma}
\end{equation}
Where $\mu$ and $\sigma$ are the mean and standard deviation over all values in the dataset.

\subsection{Weight matrix}

Local indicators of spatial association require a definition of the relationships between locations in the dataset. This is encapsulated in a weight matrix. Given a dataset with $n$ locations, a weight matrix is a $n \times n$ square matrix $W$ where element $W_{i,j}$ quantifies the relationship between location $i$ and location $j$. In its simplest form, a binary weight matrix takes $W_{ij}=1$ if location $i$ is a neighbor of location $j$ , and ${W_{ij}}=0$ otherwise. The definition of a "neighbor" is the decision of the analyst: the most common approach uses simple areal contiguity. For simplicity in the later calculation, we will row-normalize the weight matrix:
\begin{equation}
\forall i \in \{1, 2, \ldots, n\}, \quad \sum_{j=1}^{n} W_{ij} = 1
\end{equation}
Choosing an appropriate weight matrix is challenging and depends on the specific parameters and goals of the task at hand. Local indicators of spatial association are highly sensitive to the choice of weight matrix. 

\subsection{Moran's I}

The Moran's I statistic is a popular local indicator of spatial association, implemented in most major geospatial software packages and libraries \cite{moran1950notes, bivand2018comparing}. It measures spatial autocorrelation. Assuming a row-normalized weight matrix $W$, Moran's I is calculated as follows: 
\begin{equation}
I = \frac {\sum_i^n \sum_j^n W_{ij} \cdot z_i \cdot z_j}{n-1}
\end{equation}
A significant negative value for the Moran's I statistic indicates negative spatial autocorrelation, a significant positive value indicating positive spatial autocorrelation, and a value close to 0 indicates no spatial autocorrelation (spatial randomness). 
%\subsubsection{Local Moran's I}
The Local Moran's I statistic addresses the need for a more granular assessment of spatial autocorrelation by breaking the global spatial autocorrelation into a separate value for each location \cite{anselin1995local}. The Local Moran's I statistic for location $i$ is calculated as follows:
\begin{equation}
I_i = \frac{z_i \cdot lag_i}{n-1} = \frac{z_i \cdot \sum_j^n W_{ij} \cdot z_j}{n-1}
\end{equation}
Note the spatial lag term, $lag_i = \sum_j^n W_{ij} \cdot z_j$. Spatial lag is a useful concept when interpreting local spatial autocorrelation; it is essentially the weighted mean of a location's neighbors. 

A significant negative value for Local Moran's I indicates that the location is a spatial outlier (significantly different from its neighbors), a significant positive value indicates that the location belongs to a spatial cluster (significantly similar to its neighbors), and a non-significant value indicates that the location does not exhibit significant local spatial autocorrelation. Results can be further categorized by looking at how the value and spatial lag of a location compare to the mean, which is easy to do with z-score normalized values because the mean is equal to 0. The possible assignments are "high-high" (if the value and lag are both positive), "low-low" (if the value and lag are both negative), "high-low" (if the value is positive and the lag negative), and "low-high" (if the value is negative and the lag positive). 

\subsection{Geary's C}

Like Moran's I, Geary's C is a statistic which measures spatial autocorrelation but the two methods differ in their approach: Moran's I measures spatial autocorrelation using the correlation between neighboring values whereas Geary's C measures spatial autocorrelation using the square differences between neighboring values \cite{ord1995local, getis1992analysis}. Geary's C has less widespread support in geospatial software than Moran's I. Geary's C is calculated as follows: 
\begin{equation}
    C = \frac{ \sum_i^n \sum_j^n W_{ij}\cdot z^2}{2n}
\end{equation}
Geary's C takes values 0 or greater. The values are interpreted by their proximity to 1, with values less than 1 indicating positive spatial autocorrelation, and values greater than 1 indicating negative spatial autocorrelation. Values close to 1 indicate spatial randomness. 

Like for Moran's I, there is a local equivalent of the Geary's C to provide a more granular assessment of spatial autocorrelation \cite{anselin2019local}. Local Geary's C is calculated as follows: 
\begin{equation}
    C_i = \sum_j^n w_{ij} \cdot (z_i - z_j)^2
\end{equation}
In essence, the Local Geary's statistic is a weighted sum of the squared difference between a location's value and its neighboring values. The statistic takes values 0 or greater. Unlike for the global Geary's C, the value of the local Geary's C statistic has no inherent meaning --- the "no spatial autocorrelation" point is no longer equal to 1. Instead, it must be interpreted with significance testing: values that are significantly lower than expected indicate positive spatial autocorrelation, values significantly higher than expected indicate negative spatial autocorrelation, and values not significantly different than expected indicate spatial randomness. Like for Local Moran's, a Local Geary's C result can be further specified by inspecting the location's value and spatial lag. However, unlike for Local Moran's I this is not always possible. If the Local Geary's C statistic indicates positive spatial autocorrelation, then the following assignments can be made: "high-high" (if the value and lag are both positive), "low-low" (if the value and lag are both negative), and "other positive spatial autocorrelation" (the remaining cases). If the Local Geary's C statistic indicates "negative spatial autocorrelation" then this result can't be specified any further. 

\subsection{Getis-Ord G}

The Getis-Ord G family of statistics differ from Moran's I and Geary's C in that they do not measure spatial autocorrelation, but instead directly measure hot-spots and cold-spots \cite{getis1992analysis}. A hot-spot is a group of neighboring locations with significantly higher than expected values, whereas a cold spot if a group of neighboring locations with significantly lower than expected values. The Getis-Ord General G statistic is a global measure that indicates whether a geospatial dataset exhibits clustering overall and whether that clustering is generally of high values or low values. The Getis-Ord General G statistic is calculated as follows: 
\begin{equation}
    G = \frac{\sum_{i=1}^n \sum_{j=1}^n W_{ij} \cdot z_i \cdot z_j}{\sum_{i=1}^n \sum_{j=1}^n  \cdot z_i \cdot z_j}, \ \ \text{where} \ j \neq i
\end{equation}
Getis-Ord General G must be interpreted in relation to a reference distribution (usually obtained using permutation testing, see \cref{sec:significance}). If the observed value of G is significantly higher than the expected value then the dataset exhibits overall clustering of high values (hot-spots), if the observed value of G is significantly lower than the expected value then the dataset exhibits overall clustering of low values (cold-spots)

The Getis-Ord Gi* and Getis-Ord Gi statistics are two local statistics which indicate whether a specific location belongs to a hot-spot or cold-spot. The Getis-Ord Gi* statistic is calculated as follows: 
\begin{equation}
    G_i^* = \frac{\sum_{j = 1}^n W_{ij} \cdot z_j}{\sqrt{\frac{1}{n-1} \cdot [n \cdot(\sum_{j=1}^n W_{ij} ^2)-1]}}
\end{equation}
The Getis-Ord $G_i$ statistic is similar, except it doesn't include the value at the focal location: 
\begin{equation}
    G_i = \frac{\sum_{j = 1}^n W_{ij} \cdot z_j - \bar{z}(i)}{S(i) \cdot \sqrt{\frac{1}{n-1} \cdot [n \cdot(\sum_{j=1}^n W_{ij} ^2)-1]}}
\end{equation}
Where $\bar{z}(i)$ and $S(i)$ are the mean and variance over all z-normalized values excluding the value at location $i$. The interpretation of Getis-Ord Gi and Gi* is similar. If the observed value of G/G* is significantly less than 0 then that location exhibits a clustering of high values (a hot-spot). If the observed value of G/G* is significantly greater than 0 then that location exhibits a clustering of low values (a cold-spot). If the observed value is not significantly different from the expected value, this indicates that there is no spatial clustering of values. 

\subsection{Assessing significance}
\label{sec:significance}

In order to correctly interpret indicators of spatial association, it is necessary to determine whether or not a value is significant. The standard approach is to use a permutation test because, unlike an analytical derivation of the statistic's theoretical distribution, it avoids imposing unrealistic assumptions upon the data \cite{anselin1995local}. A permutation test on a spatial statistic involves shuffling the data values across the spatial locations a number of times, calculating the statistic for each shuffle, and then using those values to build an empirical distribution against which the actual value can be tested. This approach differs slightly for global and local spatial statistics: for global statistics, all values are shuffled, for local statistics, the focal location's value is fixed and all other values are shuffled. A pseudo p-value is calculated by first counting the number of permuted values of the statistic which are more extreme than than the actual value:
\begin{equation}
R = \min(|\{s \in S \mid s > v \}|, M - min(|\{s \in S \mid s > v \}|)
\end{equation}
Where $M$ is the number of permutations performed, $S$ is the set of permuted values for the statistic, and $v$ is the actual value of the statistic on the dataset. The pseudo p-value can now be calculated from $R$ and $M$: 
\begin{equation}
    p* = \frac{R+1}{M+1}
\end{equation}
The interpretation of a pseudo p-value differs somewhat from the interpretation of a traditional, analytical p-value. It is important to remember that the value of $p*$ depends on the number of permutations performed. The usual p-value cutoffs (e.g. p < 0.05, p < 0.01) are often used when interpreting pseudo p-values, but the analyst should be aware of the dependence on $M$ and avoid thinking of pseudo p-values as equivalent to traditional p-values.  For this reason, we use $p*$ to refer to pseudo p-values in this manuscript and within the ClusterRadar web-tool. 

To get the upper and lower significance boundaries, we can re-arrange the $p*$ equation and set $p*$ to a specific p-value cut-off threshold $p* = p_{cutoff}$, giving:
\begin{equation}
 R_{cutoff} = \lfloor p_{cutoff} \cdot (M+1) - 1 \rfloor 
\end{equation}
To apply this index, we need $S'$, a numerically sorted list of the permuted values $S$. With $S'$ and $R_{cutoff}$, the calculation for the upper and lower cut-off values for the statistic are:
\begin{equation}
    s_{lower} = S'_{R_{cutoff}} 
\end{equation}
\begin{equation}
    s_{upper} = S'_{M - R_{cutoff}}
\end{equation}

In order to simplify the interpretation of the spatial statistics in the ClusterRadar tool, the values are z-score normalized with the permuted set of values $S$. 

\section{The ClusterRadar web-tool}
\label{sec:tool}

ClusterRadar is a web-tool that allows users to perform spatial cluster analysis and inspect the results in an interactive dashboard. The tool is fully in-browser and executes entirely on the client side. ClusterRadar performs spatial clustering using several popular local indicators of spatial association: Local Moran's I, Local Geary's C, Getis-Ord Gi, and Getis-Ord Gi*. By default, the  tool will perform only one of the Getis-Ord methods (Gi*) due to their similarity, but the user is free to enable and disable methods to suit their needs. ClusterRadar operates on areal geospatial data. The dashboard (see \cref{fig:teaser}) consists of five plot panels, each providing a different perspective on the results. Each panel is interactive, allowing the user to gain more details about a specific element through mouse events such as hovering, clicking, and dragging. The panels are interactively linked: when the user interacts with one panel that interaction is reflected in the other panels. In addition to the plot panels, ClusterRadar has a tool bar with options for uploading and configuring data, enabling and disabling methods, switching coloring modes, and downloading the data. To encourage use among a diverse set of users, ClusterRadar also features a short tutorial which describes the tool's purpose and highlights its features. 

\subsection{Coloring}
\label{sec:coloring}

\begin{figure}[bt]
  \centering 
  \includegraphics[width=\columnwidth, alt={A figure showing how the aggregate color scheme works using several examples.}]{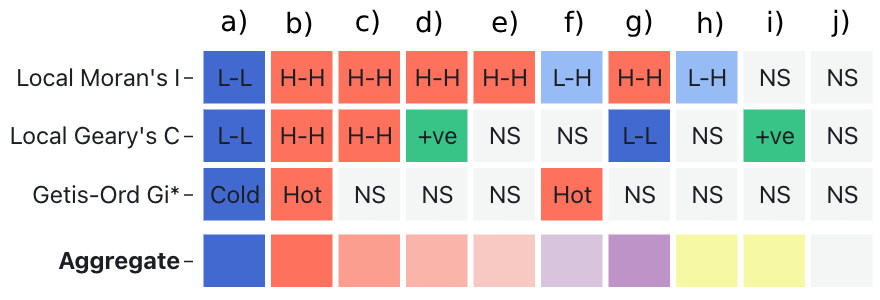}
  \caption{%
  Examples of how colors are assigned using the individual and aggregate color schemes. This plot shows examples of how the individual cluster assignments are aggregated into a single aggregate color. The aggregate color scheme visually represents the level of agreement between clustering methods. Locations are assigned a core color: red for "high-high" or "hot-spot", blue for "low-low" or "cold-spot". The final color is adjusted based on the presence of non-conflicting assignments: assignments that don't match the core assignment but don't directly contradict it either. A purple color is used for conflicts: a deeper purple for more significant conflicts. A yellow color is used for other assignments that do not fall naturally into the high or low cluster groups. Examples in this plot include total agreement in the main positive groups (a and b), partial agreement (c, d, e), minor conflict (f), major conflict (g), other (h,i) and not-significant (j).  
  }
  \label{fig:agg_color}
\end{figure}

The primary cluster map panel, zoomed map reel panel, and cluster assignments cell plot panels (see \cref{fig:teaser}a, c, and e) are all colored according to the same scheme. The tool supports several coloring modes: one for each local indicator and an aggregate mode that summarizes the assignments across all enabled methods. The single indicator modes employ a simple categorical color scheme wherein each location is colored according to the assignment from that method. As discussed in \cref{sec:methods}, the set of possible assignments differs by indicator, and therefore the color scheme for each single indicator mode differs slightly. However, for coherency, similar concepts are assigned the same color across modes, e.g. a 'high-high' assignment from Local Moran's I and a 'hot-spot' assignment from Getis-Ord Gi* are both assigned a red color. We will refer to "high-high" and "hot-spot" assignments with the joint label of "high cluster", and "low-low" and "cold-spot" assignments with the joint label of "low cluster".

The aggregate color scheme attempts to show, at a glance, the extent to which the different clustering methods agree on their assignments at a location. The scheme employs linear color scales to represent agreement, with bolder, more saturated colors representing a greater degree of agreement and muted, less saturated colors representing a lesser degree of agreement. A location at a specific timestep is first assigned a core group color: red, if a high cluster assignment is present among the assignment set, or blue, if a low cluster assignment is present among the assignment set. If both are present, then the location is a assigned a deep purple color representing a major conflict. 

With the core color assigned, the final color is decided by the proportion and nature of non-conflicting contradictory assignments in the assignment set. A non-conflicting contradictory assignment is one which does not strictly fall into the same assignment group as the core assignment, but also does not necessarily contradict it. For example, if a location is assigned "high-high" by Local Moran's I and "not significant" from both Getis-Ord Gi* and Geary's C, then the core assignment group is "high cluster" and each of the "non significant" assignments are treated as non-conflicting contradictory assignments. At timestep $t$, the assigned color at location i is calculated as follows:
\begin{equation}
    \text{color}_{t,i} = \text{col}(g_{t,i}) + h_{t,i} \cdot [\text{col}(g_{t,i}) - \text{col}(\text{"not significant"})]
\end{equation}
Where $\text{col}(g_{t,i})$ returns the an RGB color space vector representing the color assignment for the core group $g_{t,i}$. The core color assignments are a red color for the "high cluster" group, a blue color for the "low cluster" group, and a light grey for the "not significant" group. The factor $h_{t,i}$ is a scalar encapsulating the degree of non-conflicting disagreement:
\begin{equation}
h_{t,i} = \frac{1}{|L|} \cdot \sum_{l \in \text{L}_{t,i}} d(g_{t,i},l)  
\end{equation}
Where $\text{L}_{t,i}$ is the set of cluster assignments for location $i$ at time step $t$, and $d(g,l)$ returns a number between 0 and 1 representing the extent of disagreement between core group $g$ and assignment $l$. The extent of disagreement is $d(g,l) = 0$  for labels which belong in the group and $d(g,l) = 1$ for non-conflicting contradictory labels of the group. There is also a special case of $d(g,l)=0.5$ when the core group is "high cluster" or "low cluster" and the the assignment $l$ is "other positive spatial autocorrelation". This assignment is from the Geary's C method when it recognizes positive spatial autocorrelation but is unable to determine its exact nature --- a quirk of the method. It does not necessarily contradict either of the positive core group assignments, but it is not a confident agreement either, hence our decision to assign it "partial" agreement. 

There are three special cases. The first is the aforementioned major conflict case where two of the assigned labels at a location belong to conflicting groups --- this is assigned a deep purple. Sometimes assignments occur which, depending on how the methods are interpreted, can be considered a less serious conflict. For example, if Local Moran's I assigns "low-high" and Getis-Ord Gi* assigns "hot-spot", then both assignments recognize a potential positive clustering of high valued neighbors, but there is a potential disagreement on the overall assignment. The third special case is when only miscellaneous assignments and "not-significant" assignments are found in the assigned labels. For a graphical explanation of the aggregate color scheme, see \cref{fig:agg_color}.

\subsection{Primary cluster map panel}

The most prominent panel in the ClusterRadar dashboard is the primary cluster map panel: a categorical choropleth plot that colors each location by its cluster assignments (see \cref{fig:teaser}a). The choropleth plot is interactive: when the user hovers their cursor over a location, that location is brought into focus until they move the cursor elsewhere. The statistical density plot, cluster assignments cell plot, and statistical time-series panels are updated to show information pertaining to the in-focus location; see the individual descriptions of these panels for more details. The user can also keep a location in focus by clicking on it, meaning it will not be taken out of focus when their cursor leaves that location. When a user hovers their cursor over a location, a graphical tooltip appears showing additional information about that location (see \cref{sec:tooltip}). By default, the primary cluster map panel shows data from the most recent time point but the user can look through different time points using the time slider above the plot. Finally, when the user clicks and drags on the map, a selection box will appear allowing them to select multiple locations at once for further inspection in the zoomed cluster reel panel. 

\subsection{Zoomed cluster reel panel}

The zoomed cluster reel panel (see \cref{fig:teaser}c) allows the user to directly inspect how cluster assignments have changed over time in a selected sub-region of space. When the user has selected a sub-region using the primary cluster map panel, this panel will show a vertically stacked "reel" of choropleth plots --- one for each time step in the dataset. The user can then scroll through this reel to see the evolution of cluster assignments over time. Each of the choropleth plots in this panel have the same interactive features as the primary choropleth, excluding the ability to select the zoomed sub-region of space.

\subsection{Density plot panel}

\begin{figure}[tbp]
  \centering
  \includegraphics[width=\columnwidth, alt={A screenshot showing an example of the  density plot panel.}]{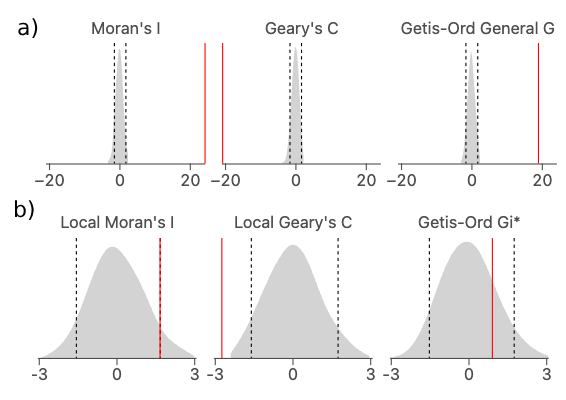}
  \caption{Examples of the density plot panel. The content of the density plot panel depends on whether a specific location is in focus. If so, the panel shows the local indicators (b), otherwise it shows the global indicators (a). Each density plot shows the empirical distribution of its respective statistic in light grey, the upper and lower significance cut-offs as dashed grey lines, and the statistic's actual value as a solid red line. If the red line is outside the dashed lines, then the statistic's value is significant. In the examples here, all global statistics are significant, and the Local Geary's C statistic is also significant. The Local Moran's I statistic is close to significant.}
  \label{fig:density}
\end{figure}

The density plot panel consists of density plots showing the empirical distributions of the enabled spatial statistics (see \cref{fig:density}), allowing the user to gain a better understanding of each statistic's distribution and significance assignments. The empirical distribution is calculated using the permutation approach described in \cref{sec:significance}. If the aggregate coloring mode is selected, then a density plot is shown for each enabled statistic, otherwise a single density plot is shown with the corresponding statistic of the selected single indicator mode. The content of the density plot panel also depends on whether or not a specific location is in focus. If so, the panel will show information regarding the local statistic(s) at that location. Otherwise, the panel will show the global statistic(s) for the whole dataset.  Each density plot consists of a filled area representation of the statistic's empirical distribution, a solid red line indicating the statistic's value, and two dashed grey lines indicating the upper and lower significance boundaries at the user's chosen significance cut-off (0.05 by default). The filled area is generated using kernel-density estimation over the permuted values. % TODO: Expand a little or refer to documentation?

\subsection{Cluster assignments cell plot panel}

\begin{figure}[tb]
  \centering 
  \includegraphics[width=\columnwidth, alt={A screenshot showing an example of the cluster assignments cell plot panel.}]{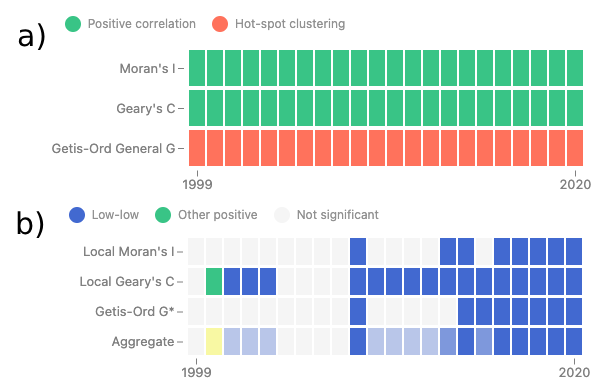}
  \caption{%
 Examples of the cluster assignments cell plot panel. The content of this panel depends on whether a specific location is in focus. If no location is in focus, the panel shows a) the global cluster assignments over time. If a location is in focus, the panel shows b) the local cluster assignments over time. In the latter case, an additional row is shown at the bottom representing the color assignments from the aggregate color scheme.
  }
  \label{fig:cell}
\end{figure}

The cluster assignments cell plot panel shows a cell plot with time on the x-axis and spatial clustering method on the y-axis (see \cref{fig:cell}). Each cell is colored according to the cluster assignment for that method at that timestep. If no location is in focus, then the cell plot shows the assignments from the global statistics. If a location is in focus, then the cell plot shows the assignments from the local statistics for that location. In the aggregate coloring mode, an additional row is added to the bottom of the cell plot which shows the aggregate color assignments. This helps the user understand how the aggregate color scheme works. The cell plot provides the most complete summary of the multi-method cluster assignments over time, but it is limited to a single location at a time.

\subsection{Statistical time-series panel}

\begin{figure}[tbp]
  \centering
    \includegraphics[width=\columnwidth, alt={A screenshot showing an example of the statistical time-series panel.}]{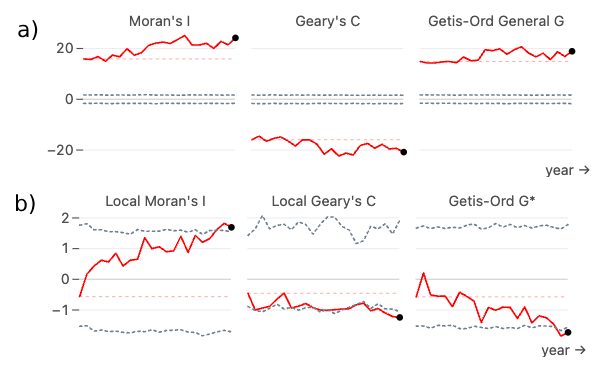}
  \caption{Examples of the time-series plot panel. The content of the panel depends on whether a specific location is in focus. If so, the panel shows b) the global statistics, otherwise it shows a) the local statistics. Each time-series plot shows the value of the respective statistic over-time as a solid red line, with the currently visualized time point marked with a black dot. A dashed horizontal pink line shows the first value as a reference. The dashed grey lines show the upper and lower significance boundaries.}
  \label{fig:time-series}
\end{figure}

The statistical time-series panel allows the user to inspect the spatial statistics over time (see \cref{fig:time-series}). If the aggregate viewing mode is selected, the time-series plot is split into sub-plots for each enabled method. Otherwise, a single plot for the current mode's method is shown. Within each plot there is a solid red line tracking the statistic's actual value over time and two dashed dark-grey lines tracking the p-value cutoffs over time. As with the cluster assignments cell plot and statistical density plot panels, the statistic(s) represented in the time-series plot depends on the current interaction state. If a specific location is in focus, then the plot will show the enabled local statistic(s) for that location. If no location is in focus, then the plot will show the enabled global statistic(s) for the entire dataset.

\subsection{Graphical tooltip}
\label{sec:tooltip}

\begin{figure}[tb]
  \centering 
  \includegraphics[width=0.6\columnwidth, alt={A screenshot showing an example of the graphical tooltip.}]{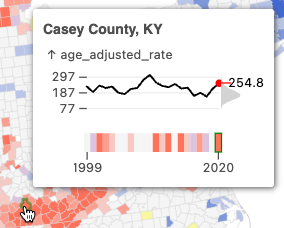}
  \caption{%
  The graphical tooltip, showing an example of the user hovering over a location. If the user has configured a geospatial name field, the top of the tooltip will show the locations name, otherwise it will show the location's geospatial ID. Below the name field is a time-series plot showing the location's value over time. Attached to the right axis of the time-series is a density plot showing the density of values across all locations and time points, with the current time point and location's value represented by a red line and text label. Below the density plot is a single-row cell plot showing the color assignments over time for the selected color mode. In this example, the color mode is the aggregate color scheme. 
  }
  \label{fig:tooltip}
\end{figure}

The graphical tooltip appears on the choropleth plots when the user hovers over a specific location, in either the main cluster map panel or the zoomed cluster reel panel (see \cref{fig:tooltip}). The tooltip allows the user to quickly get more information about a location. At the top of the tooltip is location's name or ID. Below that is a time-series plot showing the location's value over time. The axis is marked with the mean (over the whole dataset across all time points) and the mean plus or minus 3 standard deviations. Attached to the right axis of the time-series is a density plot showing the distribution of values across all time points. On this density plot, the current value is shown a red line labelled with the value in text. At the bottom is a single row cell plot showing the color assignments over-time for the in-focus location; the exact coloring shown depends on the enabled mode and is identical to the coloring shown on the map.

\subsection{Implementation details}

ClusterRadar is implemented in vanilla JavaScript and runs entirely on the client-side in the user's web browser. The basic graphical elements of the dashboard were rendered using the Observable Plot library, with interaction and additional visual elements added using D3, HTML, and CSS. Currently, there is no JavaScript library which supports the full set of spatial indicators required in ClusterRadar. The jsgeoda library supports most of them using WebAssembly but it doesn't provide estimates of the empirical distributions or significance cut-offs. Therefore, all methods were re-implemented in JavaScript. In the permutation tests, the statistics are calculated in full for each permutation --- no look-up tables or other estimation-based optimizations are used. This gives the most accurate pseudo p-values but the calculations can take a long time to run on large datasets. To ensure a responsive user interface and parallelize computation, web workers are used. When results for a given configuration are calculated, they are cached on the user's machine using IndexedDB so that the user does not need to re-run the calculations every time they visit the tool. For more details on the implementation, see the documentation and code linked in the Supplemental Materials.

\section{Usage scenario: US cancer mortality}
\label{sec:example}

\begin{figure*}[t]
  \centering 
  \includegraphics[width=\textwidth, alt={A collection of screenshots showing the ClusterRadar elements on example data.}]{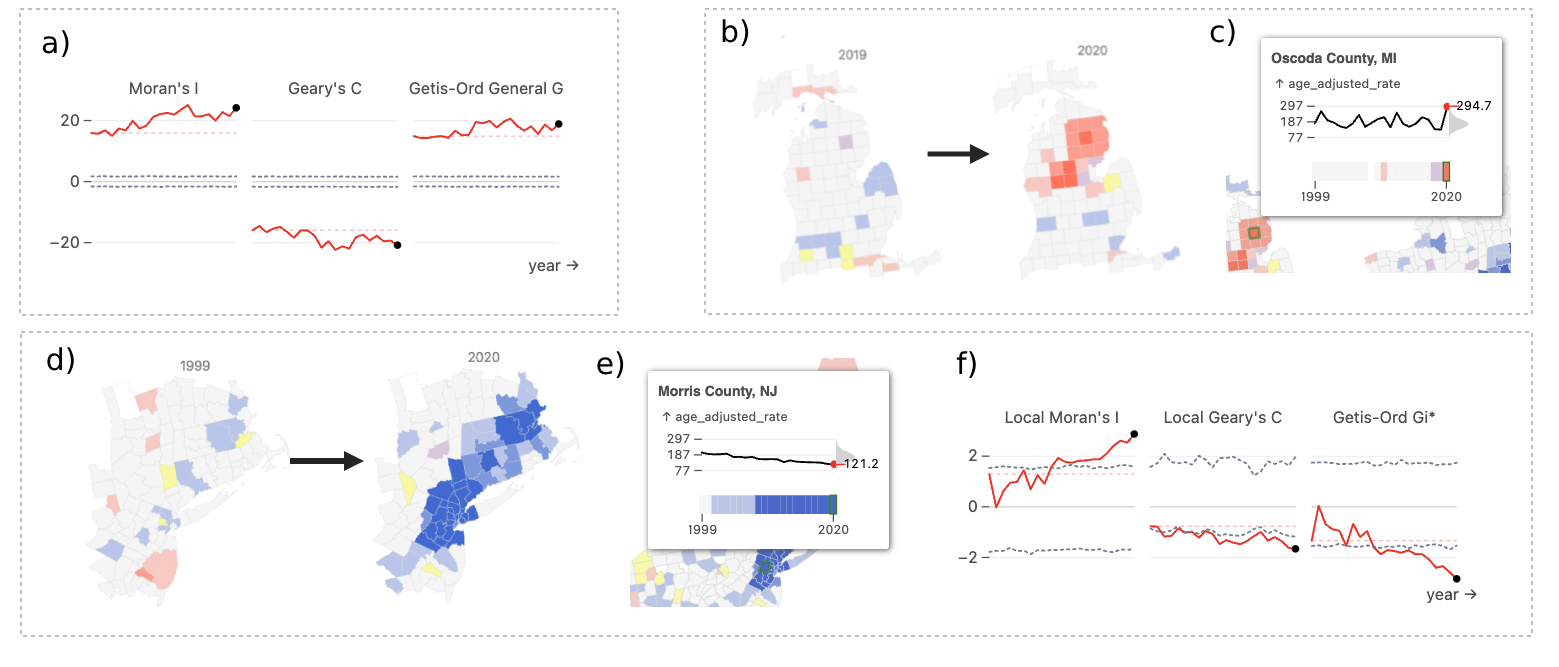}
  \caption{
  Example snapshots of ClusterRadar on a dataset of US county-level age-adjusted mortality rates, from the CDC. These illustrate some of the potential insights that ClusterRadar can help uncover. a) A snapshot of the statistical time-series panel showing the global spatial statistics. There appears to be a general increase in these statistics, especially for Moran's I. b) An edited snapshot of the cluster reel panel, showing the sudden appearance of a high cluster in Northern Michigan in 2020. c) A snapshot of the primary cluster map panel in which the user is hovering over a central area from the high cluster in Northern Michigan. The time-series shows a sudden spike in the age-adjusted mortality rate in 2020 and the cell plot shows a concomitant assignment to a high cluster. d) An edited snapshot of the cluster reel panel, showing the emergence of a low cluster in the east coast around New Jersey, New York, and Connecticut. e)  A snapshot of the primary cluster map panel in which the user is hovering over a central area (Morris County, NJ) from the low cluster on the east coast. The time-series show a gradual decline in age-adjusted mortality rates in that county. f) A snapshot of the statistical time-series panel showing the local statistics for Morris County, NJ. 
  % TODO: Either expand or contract to remove any analysis.
 }
  \label{fig:worked_example}
\end{figure*}

To show how ClusterRadar may be applied to real world data, we analyzed age-adjusted US county-level cancer mortality with a domain expert in geospatial epidemiology. The data was collected from CDC Wonder, with a filter for the "malignant neoplasms (C00-C97)" group of the IDC-10 113 cause list. The data is yearly from 1999 to 2020. We uploaded the data to the ClusterRadar tool and inspected the results in the dashboard.  We noticed the following:

\begin{itemize}
 \item An overall trend of increased global spatial autocorrelation, especially according to the Moran's I statistic (see \cref{fig:worked_example}a). The expert suggested this may be a result of overall increased spatial structure in cancer risk factors, including growing spatial disparities in smoking, obesity, and socioeconomic factors. 
 \item An emerging cold-spot in the north-eastern US around New Jersey, Pennsylvania, New York, and Connecticut (see \cref{fig:worked_example}d). The tooltip shows a steady decline in age-adjusted mortality rates since 1999, with a concomitant assignment to low clusters (see \cref{fig:worked_example}e). The statistical time-series plots reflect this this trend, with increasing local spatial autocorrelation from Local Moran's I and Geary's C, and cold-spot clustering from Getis-Ord Gi* (see \cref{fig:worked_example}f). The expert suggested that this may be caused by growing affluence in these coastal and near-coastal counties, factors associated with lower cancer mortality. Shifting demographics in these counties may be a contributing factor, though further investigation would be required to disentangle all the possible factors involved. 
 \item A large and fluctuating hot-spot in the south, around Kansas, Tennessee, and Ohio. The expert noted that her eye was drawn immediately to this hot-spot, and said that the higher rates of cancer in this region of the US are well-studied, believed to be driven by a variety of risk factors including poverty, smoking rates, and obesity. She agreed that the fluctuating nature of this hot-spot is a good example of why longitudinal analysis is important: any one time-point may present a misleading picture of confidence in the cluster's exact shape, but overall we can observe a consistent cluster of high cancer rates in that general region of the US. % TODO: Better
 \item A large cold-spot encompassing much of the western US. The expert noted that the cold spots in the coastal counties are expected, given the typically higher socioeconomic status of those counties. She also cautioned about placing too much emphasis on the other counties due to low populations  --- the cluster draws the eye due to its sheer geographic size, but encapsulates a relatively small number of people. On top of that, there is a lot of suppressed data in that part of the country due to small population sizes.
 \item A sudden hot-spot in northern Michigan in 2020 (see \cref{fig:worked_example}b). The graphical tooltip shows that this primarily results from a sudden spike in Oscoda County, MI (see \cref{fig:worked_example}c). Oscoda County has a relatively small population which can lead to unstable rates. This suggests that this particular cluster may just be noise, a result which emphasizes the importance of longitudinal analysis of spatial clusters. 
\end{itemize}
The expert emphasized that these observations are purely exploratory, and that robust statistical testing would need to be done to investigate them further. 

\section{Feedback}
% TODO: Make sure to include all bits of feedback that I reference in the discussion.
To provide an initial evaluation of ClusterRadar, we sent an evaluation survey to individuals with a diverse set of analytical interests. All survey participants have a research background, with interests spanning data science, public health, geospatial epidemiology, and computer science. The form and a snapshot of its responses at the time of writing are available in the supplemental materials. 

All participants responded that they had a basic understanding of spatial data and analysis, but only one participant claimed expert knowledge of the methods; the remaining participants expressed a mixture of unfamiliarity and basic understanding. This is a useful test base for ClusterRadar because the tool is designed for non-expert users. 

When asked if the insight provided by temporal analysis of spatial clusters is worth the additional complexity, 4 of 6 (66\%) of participants answered affirmatively, and one  participant answered that it is potentially useful but may not be worth the additional complexity. When asked whether the comparison of multiple clustering methods is useful, 2 of 6 (33\%) of users answered affirmatively, 3 of 6 (50\%) answered that it is potentially useful but may not be worth the additional complexity, and 1 of 6 answered they remained uncertain but enjoyed viewing the comparison across methods. 

In the evaluation of the tool itself, 3 of 6 (50\%) participants said ClusterRadar is successful in its primary goal of making the analysis of spatial clusters over time more accessible. The remaining participants said it is "somewhat" successful, with one stating that the tool is more descriptive than analytical. One participant expressed a desire for more information on the methods involved in the tool and how they differ. 

The participants were asked the extent to which the major features of ClusterRadar are useful. The primary cluster map panel and the graphical tooltip were deemed "very useful" by all 6 participants. The following features received a mixture of "very useful" and "somewhat useful" responses: the zoomed map reel panel (66\% "very useful"), the cluster assignments cell plot panel (50\% "very useful"), the statistical time-series panel (66\% "very useful"), the aggregate color scheme (66\% "very useful"). The density plot panel received 50\% "very useful", 33\% "somewhat useful", and a single "not useful" response: that participant commented that they did not know how to interpret the density plots. All participants said that the tool's implementation as a web-application was "very useful". One participant commented that the plot choices were "very insightful". 

\section{Discussion}

% TODO: The advantages and disadvantages of the longitudinal approach
% TODO: The decision to focus on areal data
% TODO: The decision to apply LISA-based clustering methods over other types
% TODO: Exploratory / hypothesis generation, p-value correction
% TODO: Mention the mixed success of the feedback more clearly (only 50% said it achieved its goal!)

In this paper, we have introduced ClusterRadar, a web-tool which confronts the challenge of multi-method, temporal exploration of spatial clusters using multi-faceted interactive visualization. In this section, we will discuss some of the problems we tackled while creating this tool.% , using the feedback to evaluate the success of our solutions.

\subsection{Visualizing large volumes of complex results}

A core challenge in multi-method, spatiotemporal cluster analysis is managing the potentially large volume of results. In our example (\cref{sec:example}), applying just 3 clustering methods across 22 time points to 3,143 counties yields over 200,000 individual results. ClusterRadar addresses this complexity through a multi-plot dashboard that follows the visual analytics principle of "overview first, details-on-demand" (D3). This approach allows users to gain a broad understanding of the results before drilling down into specific details. This philosophy is reflected in several parts of the tool, including the responsive nature of the density, cell, and time-series panels. As discussed in \cref{sec:tool}, each of these visualizations initially represent global statistics but changes to represent local statistics when the user interacts with a specific location. The global statistics provide the "overview", and the local statistics provide the "details-on-demand". This approach is well supported by the visual analytics literature, and ClusterRadar follows established principles to simplify the difficulties which may arise when exploring multiple visualizations at once, including the principle of maintaining the user's mental map between plots using linked interaction \cite{cui2019visual}. Feedback from a varied set of users regarding the plots was generally positive. However, some plots (e.g. the primary cluster map) received greater feedback than others (e.g. the statistical density plots), so a future effort to provide alternatives and collect feedback may benefit the tool's overall success in representing complex results. 

%The feedback regarding this approach was largely positive, with partic

\subsection{Visualizing multi-variate results over space}

The results generated by ClusterRadar are both spatial and multi-variate/multi-method (D2). Visualizing the spatial distribution of multi-variate results is a well-documented challenge in the geospatial visualization literature and a broad range of solutions have been proposed. Once again, we took a visual analytics approach, with the aggregate color scheme offering the overview, and user interaction providing further details.  The aggregate color scheme provides a way for the user to inspect the level of agreement between results at a glance. While some finer distinctions between methods might be obscured in this simplification, it provides a valuable starting point. For users wishing to delve deeper into the nuances, the interactive, linked views provide the necessary granularity to dissect and compare results with greater precision. The feedback from a varied set of users was generally positive, with some expressing confusion over certain aspects of the color scheme. Future work should tackle this confusion, perhaps with a further simplified color scheme or more detailed documentation. 

% The aggregate color scheme provides a way for the user to inspect the level of agreement between results at a glance. We designed the color scheme to consider the types of agreement and disagreement that would be most important to an analyst, without overloading them with information. Specifically, the color scheme focuses on representing hot spots and cold spots, with more nuanced results categorized as "other". While some finer distinctions between methods might be obscured in this simplification, it provides a valuable starting point. For users wishing to delve deeper into the nuances, the interactive, linked views provide the necessary granularity to dissect and compare results with precision. The feedback from a varied set of users was generally positive, with some expressing confusion over certain aspects of the scheme. Future work should tackle this confusion, perhaps with a further simplified color scheme or more detailed documentation. 

\subsection{Visualizing temporal results over space}

A challenging element of the ClusterRadar results is that they are both spatial and temporal (D1) --- elements that can be difficult to visualize together. Like with multi-variate spatial data, the problem of representing spatiotemporal data in plots is well documented in the visualization literature. There are two key approaches: contrasting time and contrasting space \cite{pena2019comparison}. In the former, some representation of the temporal results (e.g. a time-series) is plotted at each geospatial location, and in the latter separate geospatial plots are used to represent data from different time-points. The contrasting time approach can be overwhelming when there are a lot of geospatial locations and the temporal results are complex, as is the case in ClusterRadar, and so we decided on the contrasting space approach. This was achieved using the cluster map reel panel,  which stacks choropleths from different time points on top of each other, and the interactive time-slider in the primary cluster map panel, which allows users to "animate" the cluster results by viewing data from different time-points. The statistical time-series panel, cluster assignments cell plot panel, and graphical tooltip all provide additional information on temporal elements of the results. Feedback for this approach was positive, with all of the aforementioned elements receiving a positive response from survey participants. 

\subsection{Usability to non-expert users}

A key element in the success of novel applications is their usability and accessibility to the target user base. In this case, our target user base is anybody interested in the analysis of spatial clusters. While a working knowledge of spatial data, statistical principles, and common visualizations is beneficial, the ClusterRadar tool is designed to minimize the required expertise. It does so in a few ways: applying a varied set of visualizations to streamline the interpretation of the results (D3), employing a goal-focused design philosophy (D4), and distributing the application on the web (D5). Importantly, the user can access the application from any modern browser without then need for installation. ClusterRadar eliminates the need for specialized technical knowledge of the methods; users simply upload data, configure settings, and receive results. The positive feedback from users of varying backgrounds suggests the success of this approach, though additional work is required to address certain confusing elements. Several individuals expressed confusion over aspects of the tool which could potentially be cleared up through better documentation and a refined tutorial. 

\subsection{Exploratory analysis}

When inspecting the cancer mortality data in \cref{sec:example}, the expert analyst expressed the opinion that ClusterRadar is primarily an exploratory tool, rather than an analytical one. The distinction between these concepts is important: exploratory tools can uncover patterns worthy of further analysis (hypothesis generation), but are not necessarily appropriate for more rigorous inspection (hypothesis testing). Exploratory data analysis is becoming common across disciplines, including those with well-established analytical traditions such as epidemiology, driving a need for exploratory tools such as ClusterRadar \cite{gelman2021most,platt2022invited}. ClusterRadar allows the user to download the results, which an experienced analyst could then inspect in the analytical environment of their choice (e.g. R). Further work could improve the analytical appeal of ClusterRadar by supporting p-value correction or other techniques important in confirmatory spatial cluster analysis. 

% While the study participants have a broad range of backgrounds and skill-sets, they are all health researchers. Future evaluations should include participants from a  to wider domains, industry 

% TODO: Weave in more stuff about feedback, and also include how feedback only covers reseachers and not people from industry, etc. 

\section{Conclusion}

ClusterRadar advances the analysis of complex spatial datasets by offering a user-friendly environment specifically designed for interpreting temporal and multi-method spatial clustering results. By combining diverse visual elements with linked interaction, ClusterRadar tackles the challenges of representing multi-variate, spatiotemporal data. The tool's fully in-browser implementation and goal-driven design democratize spatial cluster analysis, empowering users across disciplines to uncover spatial patterns within their data. Feedback from a diverse set of users validates the utility of our approach while highlighting opportunities for further refinement.

% \section*{Availability}

\section*{Supplemental Materials}
The source code, documentation, and full survey responses are available at \url{https://github.com/episphere/ClusterRadar}. A short video outlining the core features of the tool is available at \url{https://youtu.be/mPxj1zWg47g}

%% if specified like this the section will be omitted in review mode
\acknowledgments{%
    The authors wish to thank Rena Jones, Praful Bhawsar, Jeya Balasubramanian, Meredith Shiels, Wayne Lawrence, and Rebecca Troisi for their help evaluating the ClusterRadar tool.
  Research reported in this publication was supported by the National Cancer Institute of the National Institutes of Health (CAS 10901).%
}

\bibliographystyle{abbrv-doi-hyperref-narrow}

\bibliography{references}

% \appendix % You can use the `hideappendix` class option to skip everything after \appendix

% \section{About Appendices}

% \section{Troubleshooting}

% \subsection{ifpdf error}

\end{document}